\pdfoutput=1
\documentclass[%
 aip,
 amsmath,amssymb,
 reprint,%
]{revtex4-1}
\usepackage{lineno,hyperref}
\usepackage{algorithm}
\usepackage{algorithmicx}
\usepackage{algpseudocode}
\usepackage{color,xcolor}

\usepackage{graphicx}
\usepackage{epstopdf}
\usepackage{dcolumn}
\usepackage{bm}

\usepackage[utf8]{inputenc}
\usepackage[T1]{fontenc}
\usepackage{mathptmx}
\usepackage{lineno,hyperref}
\usepackage{graphics}
\usepackage{amssymb}   
\usepackage{amsmath}
\usepackage[font=scriptsize]{caption}
\usepackage{hyperref}
\usepackage{footnote}

\def\BibTeX{{\rm B\kern-.05em{\sc i\kern-.025em b}\kern-.08emT\kern-.1667em\lower.7ex\hbox{E}\kern-.125emX}}

\begin{document}

\preprint{AIP/123-QED}

\title[\textit{Preprint submitted to Chinese Physics B.}]{A novel similarity measure for mining missing links in long-path networks}

\author{Yijun Ran}
\affiliation{College of Computer and Information Science, Southwest University, Chongqing 400715, China}

\author{Tianyu Liu}
\affiliation{College of Computer and Information Science, Southwest University, Chongqing 400715, China}

\author{Tao Jia}
\email{tjia@swu.edu.cn.}
\affiliation{College of Computer and Information Science, Southwest University, Chongqing 400715, China}

\author{Xiao-Ke Xu}
\email{xuxiaoke@foxmail.com.}
\affiliation{College of Information and Communication Engineering, Dalian Minzu University, Dalian 116600, China}

%
\begin{abstract}
Network information mining is the study of the network topology, which answers a large number of application-based questions towards the structural evolution and the function of a real system. For example, the questions can be related to how the real system evolves or how individuals interact with each other in social networks. Although the evolution of the real system may seem to be found regularly, capturing patterns on the whole process of the evolution is not trivial. Link prediction is one of the most important technologies in network information mining, which can help us understand the real system's evolution law. Link prediction aims to uncover missing links or quantify the likelihood of the emergence of nonexistent links from known network structures. Currently, widely existing methods of link prediction almost focus on short-path networks that usually have a myriad of close triangular structures. However, these algorithms on highly sparse or long-path networks have poor performance. Here, we propose a new index that is associated with the principles of Structural Equivalence and Shortest Path Length ($SESPL$) to estimate the likelihood of link existence in long-path networks. Through 548 real networks test, we find that $SESPL$ is more effective and efficient than other similarity-based predictors in long-path networks. We also exploit the performance of $SESPL$ predictor and embedding-based approaches via machine learning techniques, and the performance of $SESPL$ can achieve a gain of 44.09\% over $GraphWave$ and 7.93\% over $Node2vec$. Finally, according to the matrix of Maximal Information Coefficient ($MIC$) between all the similarity-based predictors, $SESPL$ is a new independent feature to the space of traditional similarity features.


\end{abstract}

\maketitle

\section{\label{sec:level1}Introduction}
Complex networks are widely used to represent different kinds of complex systems, in which nodes are the units of a system and links are the associations between a pair of nodes \cite{fortunato2018science,zeng2017science}. In various real systems, nodes are known but links are either missing or not present in the current frame, providing us only partial configuration of the whole network \cite{wang2011network,peixoto2019network}. For instance, genes are easy to detect in a gene regulatory network, but the interaction between genes are experimentally difficult to explore, leaving a huge gap between biologically phenomena observed and mechanisms underlying \cite{kirk2015systems}. Another case, individuals in a social network can be easily recorded, but their relationships, such as trust or distrust, like or dislike, collaborative or betrayal, are either hidden or temporal evolving \cite{girdhar2019link}. Link prediction is hence proposed to estimate likelihood of the existence of links which are either missing links that should not be, or nonexistent links that will exist in the future by utilizing currently known topology \cite{lu2011link,lu2016vital,benson2018simplicial}. While there are a myriad of factors that determine if two nodes are connected, network topology that represents the existing connectedness of a network is commonly used as the basis of link prediction \cite{clauset2008hierarchical,cannistraci2013link}. Due to potential applications, link prediction has drawn a great deal of attention during the past few years, with multiple prediction methods proposed and applied to different practical networks such as co-authorship networks \cite{wang2013quantifying,jia2017quantifying}, protein-protein interaction networks \cite{barzel2013network,kovacs2019network} and social networks \cite{ran2020generalized}. 

The widely existing methods of link prediction can be generally fallen into two categories: similarity-based approaches in network science domain and learning-based approaches introduced from the field of machine learning. The similarity-based approach is grounded in empirical evidence that the more similar two individuals (or equivalently two nodes of a network) are, the more likely that they know each other \cite{cao2019network,shang2019link}. Up to now, local similarity predictors such as Common Neighbor ($CN$), Local Path ($LP$) in similarity-based approaches are usually applied or innovated in many real networks because of their low computational complexity. For instance, L{\"u} \textit{et al.} utilized similarity-based approaches into weighted networks (\textit{e.g.,} the US air transportation network), and showed that weak ties can remarkably enhance the prediction performance \cite{lu2010link}. Soundarajan \textit{et al.} proposed a generalized common neighbor index implemented by utilizing community structure information which is added into common neighbor and indicated that the new index can improve the prediction accuracy in real-life networks \cite{soundarajan2012using}. However, these real networks in their studies are almost short-path networks. 

Because of networks with short-path, generally, have a great number of triadic closures, similarity-based approaches have highly effective performance. However, the performance of similarity-based approaches in long-path networks (i.e, a very sparse network) is poor. For example, recently Shang \textit{et al.} showed that similarity-based approaches have low prediction accuracy in tree-like networks, and then proposed the HEterogeneity Index ($HEI$) for link prediction, which performs better than many local similarity predictors \cite{shang2019link}. 

The learning-based approaches often consider link prediction as a binary classification problem to be solved by different machine learning algorithms \cite{ghasemian2020stacking}. For learning-based approaches, features are used to consist of two parts: one is the similarity features from local structures, another is derived from a representation learning such as network embedding. The network embedding technique attempts to automate feature engineering by projecting nodes in a network into a relatively low-dimensional latent space, which can locally preserve node's neighborhoods \cite{cao2019network}. After obtaining features via representation learning algorithms such as $DeepWalk$ \cite{perozzi2014deepwalk}, $Node2vec$ \cite{grover2016node2vec}, and so on, different kinds of machine learning algorithms can be used to build a classifier for link prediction. For example, recently Cao \textit{et al.} systematically compared similarity-based predictors with embedding-based predictors for link prediction and studied the shortcomings of embedding-based predictors in short-path networks \cite{cao2019network}. Likewise, here we also take advantage of $SESPL$ to compare with embedding-based predictors via machine learning techniques.

To cope with these problems, we propose a new similarity-based predictor to estimate the probability of link existence between two nodes in long-path networks. The proposed index is associated with Structural Equivalence \cite{lorrain1971structural,grover2016node2vec} and Shortest Path Length \cite{Liben2007The} hypotheses ($SESPL$). The results tested on 548 real-life networks show that $SESPL$ is highly effective in long-path networks. Our results suggest that the failure of $CN$ or $LP$ is not algorithmic, but fundamental: the hypothesis that $CN$ or $LP$ is to only capture information within the path length 2 and path length 3, respectively. Finally, we show that a machine learning approach can exploit this discrepancy between $SESPL$ and embedding-based methods by a random forest classifier. The results indicate that $SESPL$ is nearly always the best approach on 548 real networks, especially in long-path networks.
  
The remainder of the paper is organized as follows. We give a brief description of the link prediction task and empirical network data in Sec. \ref{sec:level2}. In Sec. \ref{sec:level3}, we introduce classical link prediction methods and propose the $SESPL$ index. We report the main results in Sec. \ref{sec:level4}. Finally, Sec. \ref{sec:level5} is the conclusion and discussion.

\section{\label{sec:level2}Problem definition and Data description} 
\subsection{Problem definition} 
We consider an undirected simple network $G$ composed of a set of nodes $V$ and a set of links $L$, in which a node can not connect to itself (no self-loops) nor share more than one link with another node (no repeated links). In the problem of link prediction, a predictor takes some features of the network and assigns a score $S_{ab}$ to each pair of nodes $a$ and $b$, which is proportional to the chance that nodes $a$ and $b$ should be connected. Because $G$ is undirected, the score is symmetric, \emph{i.e.,} $S_{ab} = S_{ba}$. The scores for node pairs that are not currently connected are sorted in descending order and the top candidates are likely missing links.

The $L$ links of a real-life network are randomly divided into two exclusive sets: the training set $L^T$ and the probe set $L^P$. The links in $L^P$ are considered as currently missing and need to be inferred from network topology given by the links in $L^T$. In this work, we apply the typical division \cite{lu2011link} that assigns 90\% of the $L$ links to $L^T$ and the remaining 10\% to $L^P$. In order to test the performance of link prediction, another probe set $L^N$ is used as the control group of $L^P$, which is composed of randomly chosen nonexistent links usually with the same size of $L^P$. The prediction quality is measured by comparing the score of predicted links in $L^P$ and $L^N$. If out of $n$ times of independent comparisons, there are $n'$ times that the missing link in $L^P$ has a higher score than the nonexistent link in $L^N$, and $n''$ times that the missing link and the nonexistent link have the same score \cite{lu2011link}, the result of $AUC$ can be calculated as  
\begin{align}
AUC = \frac{n' + 0.5n''}{n}.
\label{equation:auc}
\end{align}

\subsection{Data description}
\begin{figure}[ht]
\begin{center}
\includegraphics[width=0.95\linewidth]{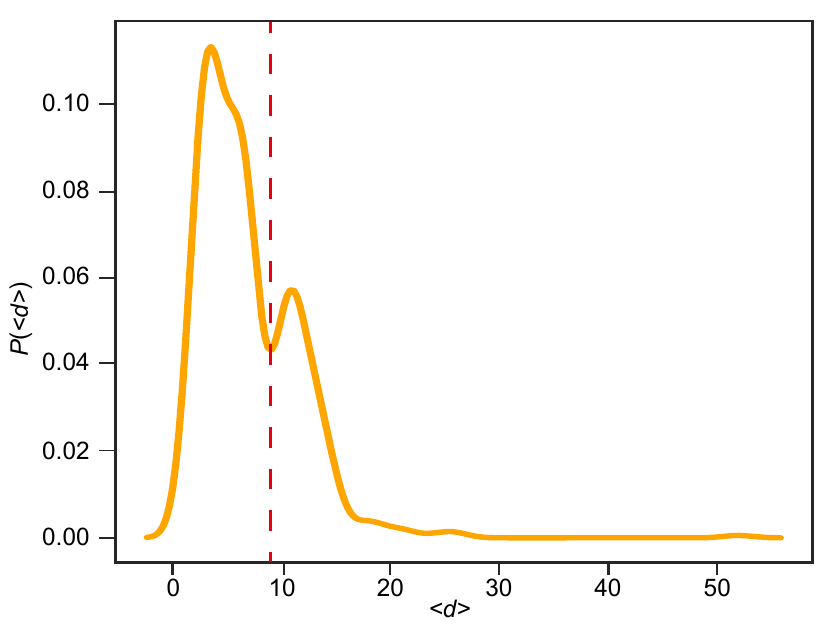}
\caption{The distribution of average shortest path length ($\langle d \rangle$) on 548 real networks. The red dash line is a turning point in which $\langle d \rangle $ is about 9. Short-path networks are located at the left of the red dash line, meanwhile long-path networks are located at the right.}
\label{fig1_distribution}
\end{center}
\end{figure}

In this study, we compare the performance of link prediction on a large corpus of 548 real networks from the CommunityFitNet corpus \cite{ghasemian2020stacking}, where there is a comprehensive description of these real networks. This structurally diverse corpus includes biological (179, 32.66\%), social (124, 22.63\%), economic (122, 22.26\%), technological (70, 12.77\%), transportation (35, 6.39\%), and information (18, 3.28\%) networks. Here, we define $\langle d \rangle$ is the average shortest path length and count the distribution of $\langle d \rangle$ on 548 real-world networks in Fig. \ref{fig1_distribution}. 

\begin{figure*}[ht]
\begin{center}
\includegraphics[width=1.0\linewidth]{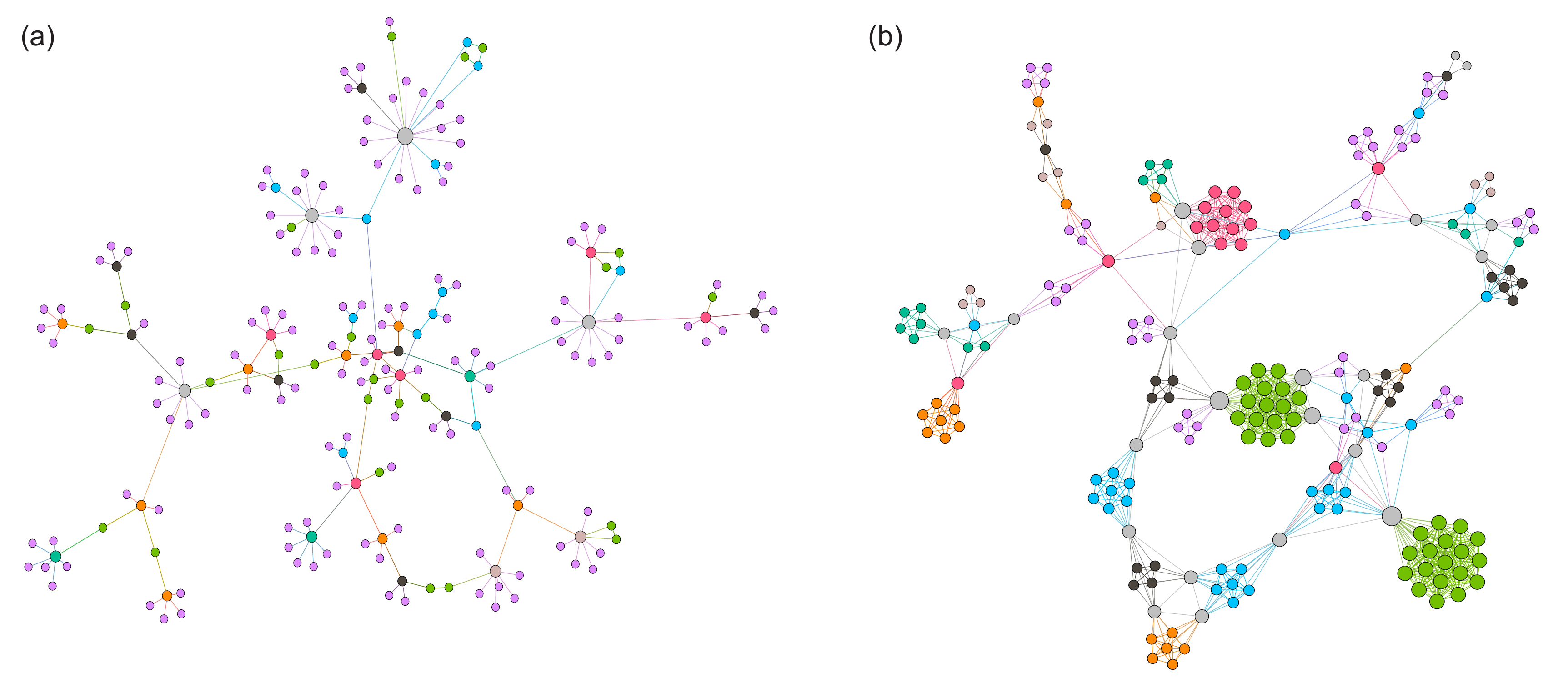}
\caption{The examples of long-path networks and short-path networks. (a) The long-path economic network has 200 nodes and 207 links, and its $ \langle d \rangle$ is 7.52. It sampled from the original economic network with $ \langle d \rangle = 10.87$ using Breadth-First Search (BFS) algorithm. (b) The short-path social network has 200 nodes and 910 links, and its $ \langle d \rangle$ is 4.73. It sampled from the original social network with $ \langle d \rangle = 6.31$ utilizing BFS algorithm. The size and color of nodes are determined according to the degree of each node.}
\label{fig2_networks}
\end{center}
\end{figure*}

In line with the distribution of $\langle d \rangle$, we define that the networks with $ \langle d \rangle < 9$ are short-path networks, otherwise the networks with $ \langle d \rangle \geq 9$ are called long-path networks. In 548 real networks, there are 164 long-path networks that primarily contain biological, technological, transportation, and economic networks. Here, we visualize one of long-path networks and one of short-path networks in Fig. \ref{fig2_networks}. The long-path network almost form a chain-like (or tree-like) network which has many open triangular structures in Fig. \ref{fig2_networks}(a). In contrast, the short-path network, has a myriad of close triangular structures (high clustering coefficient) in Fig. \ref{fig2_networks}(b).

\noindent

\section{\label{sec:level3}Methods of link prediction} 

Here, we describe in detail two categories of link predictors: similarity-based predictors and embedding-based predictors. In addition, we elaborate in more detail the method of $SESPL$.

\subsection{Similarity-based predictors}
In the predictors based on structural similarity, the simplest predictor is the method of Common Neighbor ($CN$). The $CN$ predictor is first proposed by Lorrain \textit{et al.} \cite{lorrain1971structural}, then Newman used this index to study collaboration networks \cite{newman2001clustering}. The basic idea is that two nodes share the same neighborhood are likely to share other common features hence are likely to have a link. In the problem of link prediction, the pair of nodes $a$ and $b$ is assigned a score $S_{ab}$ that depends on the set of neighborhood the two nodes share. The method of $CN$ directly counts the number of common neighbors as 
\begin{align}
S_{ab}^{CN} = \lvert {n(a) \cap n(b)} \rvert, 
\end{align}
where $n(a)$ denotes the set of neighborhood nodes that node $a$ has. Some predictors based on $CN$ have been proposed one after another, such as Salton index ($Salton$), Adamic-Adar index ($AA$), Resource Allocation index ($RA$). Our work has shown that these indices can be categorized into $CN$-based predictors. Therefore, we here select $CN$ as the representative of all $CN$-based predictors.     

The predictors based on $CN$ have been widely used in different fields because of the simplicity and interpretability \cite{jakse2003local,tamura2004prospects}. However, it is difficult for $CN$-based predictors to predict missing links when the network does not have rich close triangle structures. One more accurate method is the Local Path ($LP$) index that catches up with more path information \cite{lu2009similarity}. The $LP$ index not only considers the paths between nodes $a$ and $b$ with length 2 but also further considers that with length 3. Yielding 
\begin{align}
S_{ab}^{LP} = A^2_{ab}+ \beta A^3_{ab}, 
\end{align}
where the $A^2_{ab}$ is the number of the paths with length 2 linking node $a$ and $b$, and $\beta$ is a free parameter controlling the weights of 3-order paths. 

Due to the $LP$ index only takes 3-order path information, it may have poor performance in long-path network too. To reasonably compare with $SESPL$, we further consider the index with global topological information, the $Katz$ index \cite{katz1953new}. The $Katz$ index is one of the earliest link prediction algorithm considering all paths between nodes $a$ and $b$, which directly sums the number of all the paths. The definition is
\begin{align}
S_{ab}^{Katz} = A_{ab}+ \beta A^2_{ab}+ \beta^2 A^3_{ab}+\cdots, 
\end{align}
where $A^2_{ab}$ is the number of the paths with length 2 between nodes $a$ and $b$, and $\beta$ is a free parameter controlling the path weights. Many path-based predictors have been proposed and applied to real networks. Here, we employ $LP$ and $Katz$ as the representative path-based predictors for comparison.

Another well-known method is Preferential Attachment index ($PA$) which is based on the observation that the probability of a new link between two nodes increases as the their degrees \cite{barabasi1999emergence}. This theoretical model leads to the concept of ``the rich get richer'', which generates the power-law degree distribution observed in many real networks. Hence, the probability that a new link will connect $a$ and $b$ is proportional to
\begin{align}
S_{ab}^{PA} = k(a) \times k(b), 
\end{align}
where $k(a)$ is the degree of node $a$ and $k(b)$ is the degree of node $b$.

Above the predictors are usually applied to short-path networks. In recent researches, Shang \textit{et al.} proposed a new similarity index for link prediction in tree-like networks, named HEterogeneity Index ($HEI$), which proves that the degree heterogeneity can improve prediction performance when closed triangular structure or preferential attachment is insufficient. $HEI$ is defined by 
\begin{align}
S_{ab}^{HEI} = \lvert {k(a) - k(b)} \rvert ^ \beta, 
\end{align}
where $\beta$ is a free heterogeneity exponent. In the experiments, we set $\beta$ of $LP$, $HEI$ and $Katz$ as 0.01.

\subsection{Embedding-based predictors} 
Recently, network embedding techniques that are instances of representation learning on networks have been widely applied in link prediction \cite{cai2018comprehensive,brochier2019link}. Embedding-based predictors are derived from network embedding techniques, which attempt to automate feature engineering by projecting nodes in a network into a relatively low-dimensional latent space, to locally preserve node's neighborhoods. In this study, after representing nodes in a network as vectors, we apply t-distributed Stochastic Neighbor Embedding (t-SNE) algorithm \cite{maaten2008visualizing} to reduce the vector dimensions. The methods of dimension reduction are commonly divided into linear and non-linear approaches. For instance, both Principal Component Analysis (PCA) and Linear Discriminant Analysis (LDA) can perform a linear mapping of high-dimensional data to a lower-dimensional space. While the t-SNE algorithm is a nonlinear dimensionality reduction technique for complex high-dimensional datasets. Here, we use t-SNE to reduce the embedding vector of each node into ten dimensions space. And then we apply a Hadamard product function \cite{horadam2012hadamard} to obtain the vectors of corresponding links, which will be used as features to input into a random forest classifier. In this study, we consider three popular network embedding algorithms, $DeepWalk$ \cite{perozzi2014deepwalk}, $Node2vec$ \cite{grover2016node2vec} and $GraphWave$ \cite{donnat2018learning} for comparison.

$DeepWalk$ is the pioneered work about learning latent representation of nodes in a network \cite{perozzi2014deepwalk}. The authors applied natural language processing technology into network science. $DeepWalk$ uses local information obtained from truncated random walks to learn latent representations by treating walks as the equivalent of sentences. The representation vector learned by $DeepWalk$ reflects the local structure of a node. The more common neighbors (and higher-order neighbors) between two nodes share, the shorter distance between the corresponding two vectors of nodes.  

$Node2vec$ mainly adopts homogeneity and structural equivalence to explore diverse network structural information \cite{grover2016node2vec}. For homogeneity, $Node2vec$ can learn latent representation of nodes by embedding nodes from the same network community closer together. For structural equivalence, $Node2vec$ can learn latent representations via nodes that share similar roles should have similar embedding vectors. The embedding vectors by $Node2vec$ can be constructed for link prediction and to preserve diverse types of network information, which makes prediction performance more accurate.

$GraphWave$ is a scalable unsupervised method for learning node embeddings based on structural similarity in networks \cite{donnat2018learning}. $GraphWave$ uses a novel way that treating the wavelets as probability distributions on the network. Intuitively, a node propagates an energy unit on the network and characterizes its neighbor topology based on network response to this probe. $GraphWave$ shows that the nodes with similar structures can be closely embedded together in vector spaces. $GraphWave$ is the same as $Node2vec$, it takes advantage of similar structure information to embed nodes into a low dimension space.

\subsection{The $SESPL$ predictor}
In this section, we will present in more detail the proposed $SESPL$ predictor. Obviously, the higher score computed by a similarity-based predictor means that this index is a better predictor. In $CN$-based predictors, the idea behind is that the more common friends two individuals have, the more likely that they know each other. The $CN$-based predictors and path-based predictors show high performance in real social networks which are usually high clustering coefficients, but fail to predict the existence of a link between two nodes in long-path networks. This main reason is that these predictors are not designed to capture long (i.e., high-order) path information existing in many real-life networks. While $Katz$ predictor is based on global structure information, but it is extremely computationally expensive. Meanwhile, long-path networks such as technological, transportation, and economic networks are ubiquitous in real-world systems. Therefore, it is necessary to design an efficient and semi-local algorithm for link prediction, especially for big-size and long-path networks. 

Here we take advantage of two important and common hypotheses into our work. On the one hand, the notion of functional mapping is of central importance in ``theory of categories'', a branch of modern algebra with numerous applications to algebra, topology, logic \cite{mitchell1965theory}. This means that a group (or structure) is just a category with special properties in which all the relationships are invertible in undirected networks. The nodes with same category role in a network can be mapped into a similar structure role, that is, structural equivalence \cite{grover2016node2vec,donnat2018learning}. Therefore, there is an important ground and reason to follow when we use this hypothesis to deduce whether a link exists or not. This is due to the fact that the nodes that have similar structural roles in a network should link closely together.

On the other hand, the shorter the shortest path length between two unconnected nodes is, the easier they are to form a link. This hypothesis clearly expresses the importance of physical distance in our real-lives. For instance, in a realistic transportation system, architects often consider a variety of factors to make the highway as close as possible between the existing two transportation junctions when the government is planning to build a new highway. Similarly, technicians can add a cable between the two nearest switches so that it can reach the asynchronous and efficient transmission of information. This hypothesis also follows the notion that many real-world networks are ``small worlds'' in which individuals are related through short links, such as collaboration networks, social networks, etc. 

Overall, we mainly consider both Structural Equivalence and Shortest Path Length to measure the similarity between two nodes, namely, the $SESPL$ index. Intuitively, we note that structural equivalence is often sufficient to characterize local neighborhoods accurately. Here we quantify the structural equivalence of two nodes by only taking into account the first-order and second-order neighbors. Quantifying similarities and determining isomorphisms among graphs are the fundamental problem in graph theory, with a very long history \cite{prvzulj2007biological,aliakbary2015distance}. A recent work proposed an efficient and precise algorithm for quantifying dissimilarities among graphs, which is based on quantifying differences among distance probability distributions extracted from networks \cite{schieber2017quantification}. However, this measure has a high time complexity for big-size networks. Therefore, our idea to measure the structural equivalence, $SE(a, b)$, of two local structures is to associate to the degree centrality of each node in a local structure which can represent nodes' connectivity. The degree centrality shows that a node is central if it has many links with other nodes in a network \cite{faust1997centrality}. Degree centrality of a node $i$ can be defined as
\begin{align}
dc(i) =  \frac{deg(i)}{N_{a}-1},
\label{equation:dc}
\end{align}
where $deg(i)$ is the degree of node $i$, $N_{a}$ is the number of nodes in the local structure centered on node $a$. To perform a highly precise comparison, we consider a vector $P_{a}$ in which the elements are the degree centrality of each node $i$ in the local structure. For instance, $P_{a} = \{dc(1), dc(2), \cdots, dc(N_{a})\}$. After all nodes in the local structure have been computed, we then sort $P_{a}$ in descending order. Similarly, we compute the degree centrality of each node in the local structure centered on node $b$, so as to obtain $P_{b}$. Finally we can take use of Jensen-Shannon divergence \cite{lin1991divergence} to measure the similarity of nodes $a$ and $b$, yielding
\begin{align}
SE(a, b) =  1-\mathcal{J}(P_{a}, P_{b}),
\label{equation:se}
\end{align}
where $\mathcal{J}(P_{a}, P_{b})$ is the Jensen-Shannon divergence between the local structure centered on nodes $a$ and $b$, respectively.

\begin{figure}[ht]
\begin{center}
\includegraphics[width=1.0\linewidth]{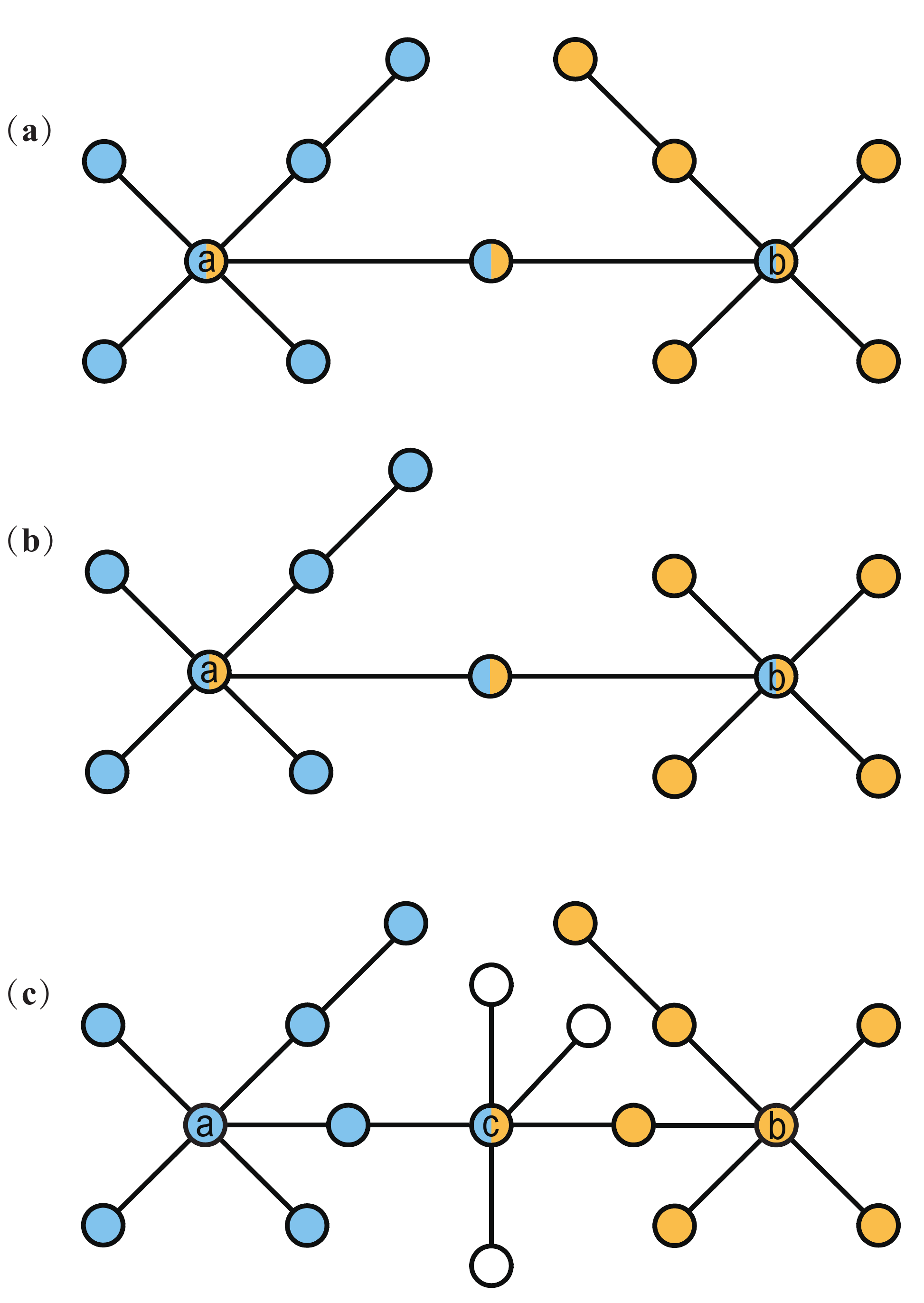}
\caption{The local structure of node $a$ or $b$ in each toy network. Different colors indicate the ownership of a node, that is to say, the blue nodes constitute the local structure of node $a$ in each toy network, the yellow nodes constitute the local structure of node $b$ in each toy network.}
\label{fig3_SESPL}
\end{center}
\end{figure}\noindent

The shortest path length hypothesis mentioned above shows that the shorter distance between two nodes is, the higher probability they have to form a link. We define $d_{ab}$ is the shortest path length of a node pair $(a, b)$ in a real-life network. The link prediction task is that predicting the chance of the existence of a link between a disconnected node pair, so $d_{ab}$ should be longer than or equal to 2. Our idea is that taking advantage of structural equivalence and the shortest path length hypotheses to facilitate the prediction. To do that, given a disconnected node pair $(a, b)$, the likelihood score of link $(a, b)$ is defined as
\begin{align}
S_{ab}^{SESPL} =  \frac{SE(a, b)}{d_{ab}-1}.
\label{equation:SESPL}
\end{align}
This $SESPL$ index means that the more similar structural role and the shorter the shortest path length between nodes $a$ and $b$ are, the higher probability to form a link. 

We provide three examples for introducing the calculation process of $SESPL$ in Fig. \ref{fig3_SESPL}. Here, we consider the first-order and second-order neighbors of a node to represent the local structure of this node. As shown in Fig. \ref{fig3_SESPL}(a), the local structures centered on nodes $a$ and $b$ which have 8 nodes (blue nodes and yellow nodes, respectively) and 7 links are equivalence, and the path between them is shortest. $P_{a} = \{ \frac{1}{7}, \frac{1}{7}, \frac{1}{7}, \frac{1}{7}, \frac{1}{7}, \frac{2}{7}, \frac{2}{7}, \frac{5}{7}\}$, and $P_{b} = \{ \frac{1}{7}, \frac{1}{7}, \frac{1}{7}, \frac{1}{7}, \frac{1}{7}, \frac{2}{7}, \frac{2}{7}, \frac{5}{7}\}$, so $SE(a, b) =  1-\mathcal{J}(P_{a}, P_{b}) = 1$. And $d_{ab}=2$, finally we get $S_{ab}^{SESPL}(n_1) =  \frac{SE(a, b)}{d_{ab}-1} = \frac{1}{1}=1$. The local structure centered on $a$ which has 8 nodes (blue nodes) and 7 links and the local structure centered on $b$ which has 7 nodes (yellow nodes) and 6 links are not equivalence, but the path between $a$ and $b$ is the shortest in Fig. \ref{fig3_SESPL}(b). $P_{a}$ stays the same, and $P_{b} = \{ 0, \frac{1}{6}, \frac{1}{6}, \frac{1}{6}, \frac{1}{6}, \frac{1}{6}, \frac{2}{6}, \frac{5}{6}\}$. Finally, $S_{ab}^{SESPL}(n_2) =  \frac{SE(a, b)}{d_{ab}-1} = \frac{0.823}{1} = 0.823$. At last, the local structure centered on $a$ and $b$ are equivalence, but the path between them is not the shortest in Fig. \ref{fig3_SESPL}(c), so $S_{ab}^{SESPL}(n_3) =  \frac{SE(a, b)}{d_{ab}-1} = \frac{1}{3}$. Taken together, $S_{ab}^{SESPL}(n_1) > S_{ab}^{SESPL}(n_2) > S_{ab}^{SESPL}(n_3)$, which supports our initial hypothesis of the $SESPL$ definition.
 
When we compute $SESPL$, there are two parts. Firstly, the time complexity is $O(Nk)$ for calculating structural equivalence if the time complexity to traverse the neighborhood of a node is simply $k$. Then, we use the Dijkstra algorithm \cite{johnson1973note} to search the path, its time complexity is $O(N^2)$. Taken together, the complexity of $SESPL$ index is roughly $O(N^2)$.

\section{\label{sec:level4}Experimental Results}
\subsection{Performance evaluation of $SESPL$} 
\begin{figure}[ht]
\begin{center}
\includegraphics[width=1.0\linewidth]{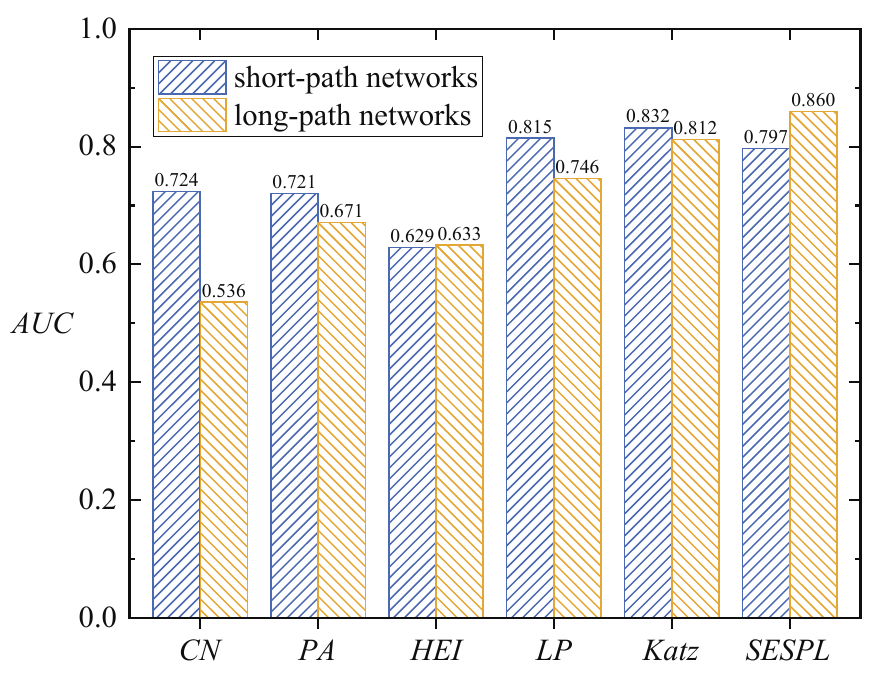}
\caption{The $AUC$ results of 6 similarity-based predictors in short-path and long-path networks.}
\label{fig4_sl_auc}
\end{center}
\end{figure}

In order to perform overall the predictive ability of $SESPL$, we quantify its performance on 548 real-world networks. Here, all the results are based on the average over $1000$ runs of simulation. Considering different properties of real-life networks, we divide them into short-path and long-path networks according to Fig. \ref{fig1_distribution}. As shown in Fig. \ref{fig4_sl_auc}, we first compare the performance of each predictor across 384 short-path networks.

\noindent 

\begin{figure}[ht]
\begin{center}
\includegraphics[width=1.0\linewidth]{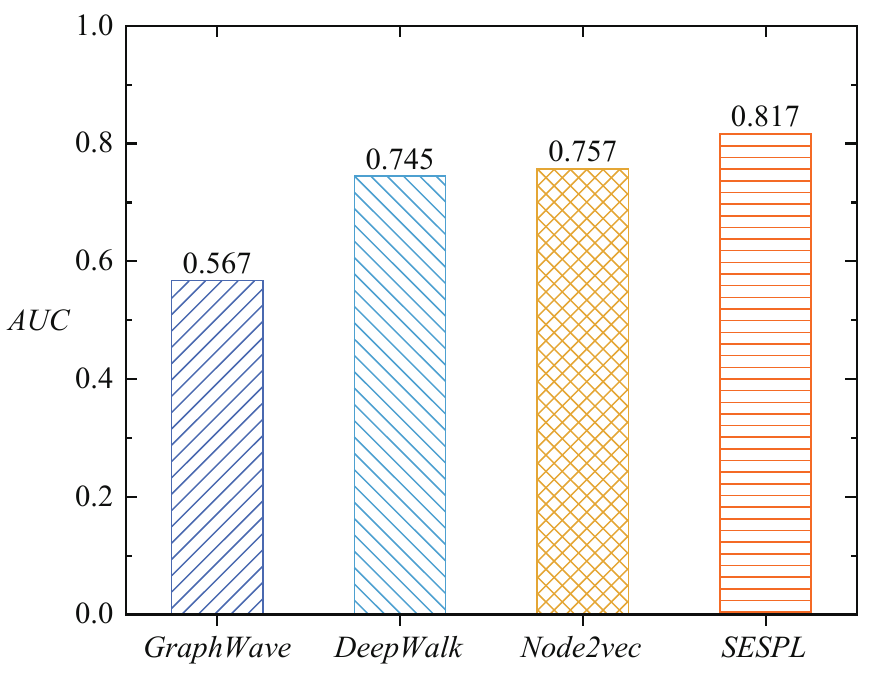}
\caption{The results of machine learning applied to 548 real networks. Each embedding vector is reduced into 10 dimensions space as 10 dimensions feature. The score of $SESPL$ as feature is inputted into a random forest classifier.}
\label{fig5_ML}
\end{center}
\end{figure}

\begin{figure*}[ht]
\begin{center}
\includegraphics[width=1.0\linewidth]{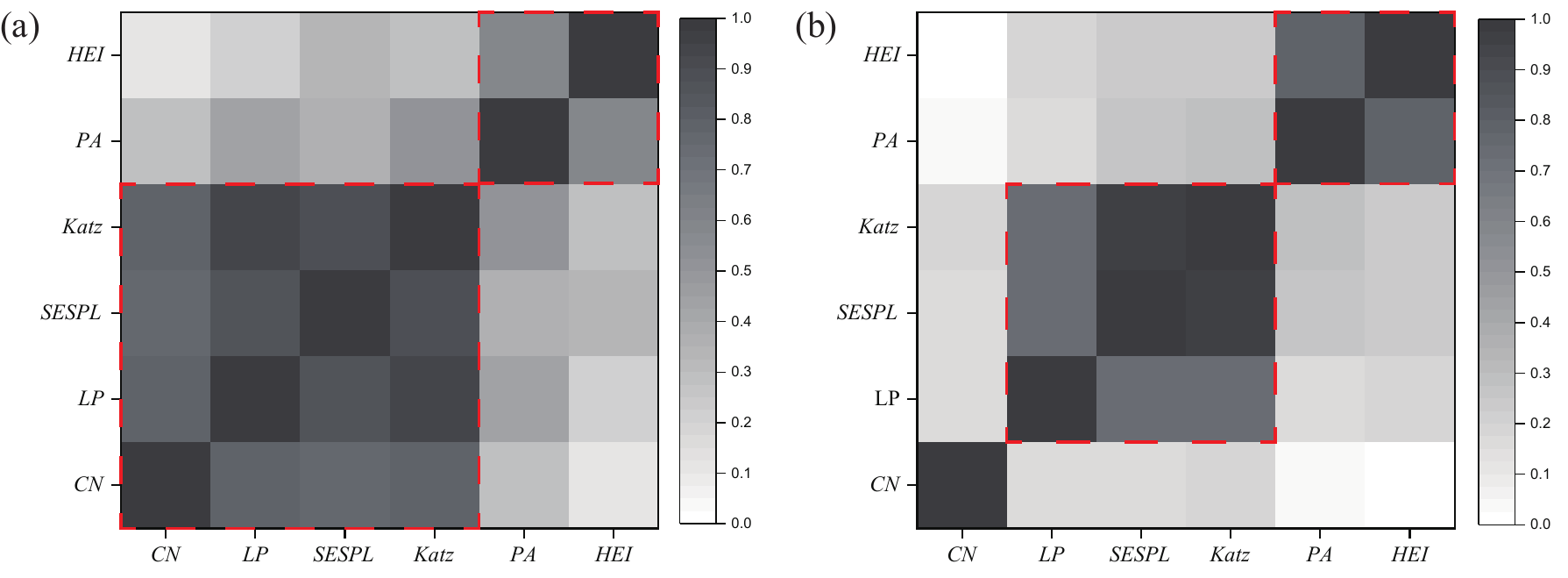}
\caption{The heat map of Maximal Information Coefficient ($MIC$) correlation matrix by the scores of links (i.e, $L^P$ and $L^N$) among six topology similarity-based features. The color intensity indicates the strength of the correlation. (a) The heat map of mean $MIC$ on 384 short-path networks. (b) The heat map of mean $MIC$ on 164 long-path networks.}
\label{fig6_MIC}
\end{center}
\end{figure*}

Although the predictive performance of $SESPL$ is higher than $CN$, $PA$ and $HEI$, it is lower than those of $LP$ or $Katz$ on 384 short-path networks. This exhibits that $SESPL$ has no advantage on short-path networks. It also verify that $LP$ is the best algorithm in short-path networks when utilizing limited resources \cite{lu2009similarity,lu2016vital}. In contrast, the performance of $SESPL$ on 164 long-path networks is shown in Fig. \ref{fig4_sl_auc}, which is higher than all the other predictors, including $LP$ or $Katz$ algorithms. The $LP$ predictor has relatively poor performance in long-path networks, because $LP$ only takes the advantage of 3-order path information. Although $Katz$ can also achieve a high performance in long-path networks, it has high computational complexity and lower performance than $SESPL$.

In general, long-path networks are more difficult to predict missing links than short path networks because the former have more sparse links. The performance of $CN$,  $PA$, $LP$ and $Katz$ algorithms support this conclusion in Fig. \ref{fig4_sl_auc}. In contrast, $SESPL$ and $HEI$ have a gain on prediction performance from short-path to long-path networks, which indicate that they are the suitable algorithms for long-path networks. Especially, $SESPL$ not only has a 7.90\% performance improvement over short-path networks in long-path networks, but also is higher than all the other algorithms in long-path networks.

Finally, to compare the difference between $SESPL$ and each embedding-based predictor, we characterize their predictive performance by training a random forest classifier. As depicted in Fig. \ref{fig5_ML}, through 548 real-world networks testing, we find that the $SESPL$ predictor significantly outperforms each embedding-based predictor. This may be because $SESPL$ can capture two kinds of network properties, that are, structural equivalence and physical distance. By contrast, $GraphWave$ has the worst predictive performance, which might be due to that $GraphWave$ only keeps structural equivalence when embedding nodes in a network into a relatively low-dimensional latent space \cite{donnat2018learning}. In addition, $Node2vec$ has relatively higher performance than $DeepWalk$, this is mainly because $Node2vec$ has a flexible neighborhood sampling strategy that can balance structural equivalence and homogeneity \cite{grover2016node2vec}. Taken together, $SESPL$ we proposed is a state-of-the-art predictor, especially in long-path networks.

\subsection{Similarity-based predictors correlation detection}
To deeply explain that $SESPL$ is the useful feature for link prediction, we utilize Maximal Information Coefficient ($MIC$) \cite{reshef2011detecting} to quantify the correlation among six similarity-based features. The correlation between two predictors $S^i$ and $S^j$ is defined as $MIC(S^i, S^j)$. The larger the $MIC(S^i, S^j)$ is, the stronger the substitutability between two features $S^i$ and $S^j$ is. $MIC(S^i, S^j)=0$ shows that $S^i$ and $S^j$ are independent of each other.

As depicted in Fig. \ref{fig6_MIC}, there is a strong correlation between predictors in each red dot-line box, while the correlation between predictors in different red dot-line boxes is weak. Overall, the correlation among $SESPL$, $CN$, $LP$ and $Katz$ is different in short-path and long-path networks, while the correlation between $PA$ and $HEI$ is the same because the two predictors take the advantage of degree information. As shown in Fig. \ref{fig6_MIC}(a), the six predictors can classify as two kinds of features. One is that $SESPL$, $CN$, $LP$ and $Katz$ can all capture common neighbor information, so the correlations among them are strong. The other is that the correlation between $PA$ and $HEI$ is strong because of utilizing degree information.

For long-path networks, however, $CN$ is almost independent of other predictors because it can not capture the high-order path information in Fig. \ref{fig6_MIC}(b). Therefore, roughly speaking, there are three kinds of features in long-path networks. The $LP$, $SESPL$ and $Katz$ predictor can be classified as the same feature. But strictly speaking, the correlation between $SESPL$ and $Katz$ is the strongest because of capturing high-order path information. That is to say, $SESPL$ and $Katz$ are the most similar feature. This shows that the $SESPL$ can totally replace the $Katz$ when caring about limited resources. Taken together, $SESPL$ can be regarded as a supplement to structure similarity features of long-path networks.

\section{\label{sec:level5}Conclusion and discussion} 
To summarize, we propose a new predictor to estimate the probability of link existence between two nodes in long-path networks, called $SESPL$, which  can capture the principles of structural equivalence and the shortest path length. $SESPL$ is highly effective and efficient compared with other similarity-based predictors in long-path networks. We also exploit the performance of $SESPL$ predictor and embedding-based approaches via machine learning techniques, and the experimental results indicate that the best prediction performance comes from $SESPL$ feature. Finally, according to the matrix of Maximal Information Coefficient (MIC) among all the predictors, the index of $SESPL$ can be regarded as a supplement to structure similarity features of long-path networks.
 
In this work, the principles of the structural equivalence and the shortest path length are integrated into similarity-based predictors, which can provide new insights into link prediction. The structural equivalence is efficiently quantified by the Jensen-Shannon divergence. In future, we will further consider higher-order network structures or other ways to quantify structural equivalence. In addition, for simplicity reasons, our index does not take link weights into consideration. The link weights, measuring how frequent two nodes are associated, is an important variable. We will also try to apply $SESPL$ predictor into link prediction of weighted networks.

\section*{Acknowledgements}
This work was supported by the National Natural Science Foundation of China (Grant No. 61773091), the LiaoNing Revitalization Talents Program (Grant No. XLYC1807106), the National Social Science Foundation of China (Grant No. 20CTQ029), the Fundamental Research Funds for the Central Universities (Grant No. SWU119062). Yijun Ran is supported by China Scholarship Council (CSC No.202006990042).

\section*{Competing financial interests} 

The authors declare no competing financial interests.

\section*{DATA AVAILABILITY}
The 548 real-world data used in this study are available from the article Amir Ghasemian, Homa Hosseinmardi, Aram Galstyan, Edoardo M Airoldi, and Aaron Clauset, `` Stacking models for nearly optimal link prediction in complex networks''. Proceedings of the National Academy of Sciences, 117(38):23393-23400, 2020. (\url{https://github.com/Aghasemian/OptimalLinkPrediction}).

\section*{References}
\bibliography{mybibfile}

\end{document}